# A note on upscaling retardation factor in hierarchical porous media with multimodal reactive mineral facies

Mohamad Reza Soltanian[1,*] Robert Ritzi[1], Chaocheng Huang[2], Zhenxue Dai[3], Hailin Deng[4]

[1]Department of Earth and Environmental Sciences, Wright State University, Dayton, OH, 45435

[2]Department of Mathematics and Statistics, Wright State University, Dayton, OH, 45435

[3]EES-16, Earth and Environmental Sciences Division, Los Alamos National Laboratory, Mailstop T003, Los Alamos, NM 87545

[4]CSIRO Land and Water, Private Bag No. 5, Wembley WA 6913, Australia

*Corresponding author. Phone: (937) 775-2201, e-mail: m.rezasoltanian@gmail.com

## Abstract

We present a model for upscaling the time-dependent effective retardation factor, $\widetilde{R}(t)$, in hierarchical porous media with multimodal reactive mineral facies. The model extends the approach by Deng et al. (2013) in which they expanded a Lagrangian-based stochastic theory presented by Rajaram (1997) in order to describe the scaling effect of $\widetilde{R}(t)$. They used a first-order linear approximation in deriving their model to make the derivation tractable. Importantly, the linear approximation is known to be valid only to variances of 0.2. In this article we show that the model can be derived with a higher-order approximation, which allows for representing variances from 0.2 to 1.0. We present the derivation, and use the resulting model to recalculate $\widetilde{R}(t)$ for the scenario examined by Deng et al. (2013).

# Introduction

Nonideal transport behavior of reactive solutes has been observed in experimental data and in numerical simulations (e.g. Roberts et al., 1986; Burr et al., 1994; Brusseau and Srivastava, 1997; Rajaram, 1997). One nonideal behavior is the temporal decrease in the average velocity of the reactive plume. This behavior causes a time-dependent effective retardation factor, $\tilde{R}(t)$, defined as the ratio of the average centroid velocity of a nonreactive plume to the average centroid velocity of a reactive plume (Brusseau and Srivastava, 1997).

Deng et al. (2013) presented a model for upscaling $\tilde{R}(t)$ in hierarchical porous media with multimodal reactive mineral facies. This model has significant practical implications in reactive transport modeling at the field scale and it provides new insight into how the effective retardation factor in porous formations is quantitatively linked to multimodal reactive mineral distributions. Deng et al. (2013) expanded on a Lagrangian-based stochastic theory developed by Rajaram (1997) to analyze $\tilde{R}(t)$. In deriving their model they assumed, like Rajaram (1997), a first order linear approximation for the perturbation of the retardation factor in order to make the derivation tractable. They illustrated their model with an example study in which the variance of the log-permeability and the log-sorption coefficient were both ~0.85.

Importantly, the linear approximation is known to be valid only to variances of 0.2 (Rajaram, 1997). In this note we show that the model can be derived with a higher-order approximation, which allows for representing variances from 0.2 to 1.0. We present the derivation, and use the resulting model to recalculate $\tilde{R}(t)$ for the scenario examined by Deng et al. (2013).

For transport of a reactive solute undergoing linear equilibrium sorption the retardation factor, $R$, is locally related to $K_d$ by the relationship $R(x) = 1 + (\rho_b/n)K_d(x)$ where $\rho_b$ and $n$ are the bulk density and porosity of the medium, respectively. Consider steady groundwater flow in a three dimensional unbounded saturated porous formation with a mean hydraulic gradient, $J$, oriented in $x_1$ direction. Deng et al. (2013) used the following Lagrangian-based expression from Rajaram (1997) to analyze $\tilde{R}(t)$:

$$\frac{\bar{R}}{\tilde{R}(t)} = (1+\frac{\sigma_R^2}{\bar{R}^2}-\frac{\sigma_{v_1 R}}{\bar{v}_1 \bar{R}}) - \frac{\tilde{R}(t)}{\bar{R}}\{\frac{1}{\bar{R}^2}[\sigma_R^2 - C_{RR}(\frac{\bar{v}_1 t}{\tilde{R}(t)},0,0)] - \frac{1}{\bar{v}_1 \bar{R}}[\sigma_{v_1 R} - C_{v_1 R}(\frac{\bar{v}_1 t}{\tilde{R}(t)},0,0)]\} \quad (1)$$

where $\bar{v}_1$ is the average groundwater velocity in the mean flow direction, $x_1$, $\bar{R}$ is the arithmetic mean of retardation factor, and $C_{RR}(\xi_1,\xi_2,\xi_3)$ is the two-point spatial covariance of $R$ (in general form with arguments $\xi_i$ being components of the lag vector, but written in equation (1) for the mean flow direction only). $C_{v_1 R}$ is the spatial cross-covariance of $v_1$ and $R$ along the mean flow direction. The $\sigma_R^2$ and $\sigma_{v_1 R}$ are the variance of $R$ and point covariance of $v_1$ and $R$, respectively. Equation (1) shows that the time-dependent behavior of $\tilde{R}(t)$ is determined by $C_{RR}$ and $C_{v_1 R}$.

**Derivation of $C_{RR}$ and $C_{v_1 R}$ using higher-order approximation**

In this note the goal is to calculate $\tilde{R}(t)$ using the Lagrangian-based expression by Rajaram (1997), and that for this purpose we need to derive the related expressions for $C_{RR}$ and $C_{v_1 R}$. In order to derive $C_{RR}$ and $C_{v_1 R}$ we use $K_d$ and the hydraulic conductivity, $K$, as random variables. We assume that these random variables are second order stationary and log normally distributed, as did Deng et al. (2013). The retardation factor can also be expressed as $R(x) = 1 + (\rho_b/n) e^{w(x)}$ where $w(x)$ is $\ln K_d(x)$. Using stochastic theory $w(x)$ can be replaced by $\bar{w} + w'$ where $\bar{w}$ and $w'$ are the mean and perturbation of $w(x)$. Deng et al. (2013) used the following first order linear approximation presented by Rajaram (1997) in order to find the perturbation of $R$ :

$$R' \cong \frac{\rho_b}{n} K_d^G w' \qquad (2)$$

where $K_d^G$ is the geometric mean of $\ln K_d(x)$. The linear approximation is limited to variance of $\ln K_d$ around 0.2 (see Appendix A in Rajaram, 1997). In the following we use a higher-order approximation for the perturbation of $R$ in order to increase the limit on variance of $\ln K_d$.

The average retardation factor, $\overline{R}$, is derived using the Taylor series expansion for $e^{w'}$ and the point that $E[w'] = 0$ as:

$$\begin{aligned}\overline{R} &= 1 + \frac{\rho_b}{n} K_d^G (E[1 + w' + \frac{w'^2}{2!} + ...]) \\ &= 1 + \frac{\rho_b}{n} K_d^G (E[1 + \frac{w'^2}{2!} + ...]) \\ &= 1 + \frac{\rho_b}{n} K_d^G e^{[\frac{\sigma_w^2}{2}]} + ... \qquad (3)\end{aligned}$$

By substitution, the perturbation of $R$ is:

$$R' = R - \overline{R} = \frac{\rho_b}{n} K_d^G (e^{w'} - e^{[\frac{\sigma_w^2}{2}]}) \qquad (4)$$

Consequently, by using the Taylor series expansion for $e^{w'}$ and the point that $E[w'] = 0$, the two-point spatial covariance of $R$ is derived as:

$$C_{RR}(\xi) = <R'(x) R'(x+\xi)> = (\frac{\rho_b}{n} K_d^G)^2 e^{[\sigma_w^2]} (e^{C_{ww}(\xi)} - 1) \qquad (5)$$

where $\xi$ is separation distance or lag distance, and $C_{ww}(\xi)$ two-point spatial covariance of $\ln K_d(x)$ which is explained below. Therefore, the corresponding variance of the retardation factor is:

$$\sigma_R^2 = (\frac{\rho_b}{n} K_d^G)^2 e^{[\sigma_w^2]}(e^{[\sigma_w^2]} - 1) \qquad (6)$$

Deng et al. (2013) considered a porous media domain $\Omega$ filled with $N$ reactive mineral assemblages (RMA) of mutually exclusive occurrences. Let $Y(x)$ be multimodal spatial random variables for $\ln K$ or $\ln K_d$ at location $x$. It can be expressed using indicator geostatistics as:

$$Y(x) = \sum_{j=1}^{N} I_j(x) Y_j(x) \qquad (7)$$

where $I_j(x)$ is indictor variable within the domain $\Omega$ and $Y_j(x)$ are variables of the $j$-th RMA. Following Ritzi et al. (2004), the composite mean $M_Y$ and variance $\sigma_Y^2$ of $Y_j(x)$ are computed as (see also Soltanian et al., 2014; Soltanian and Ritzi, 2014):

$$M_Y = \sum_{j=1}^{N} p_j m_j \qquad (8)$$

$$\sigma_Y^2 = \sum_{j=1}^{N} p_j \sigma_j^2 + \frac{1}{2} \sum_{i=1}^{N} \sum_{j=1}^{N} p_j p_i (m_i - m_j)^2 \qquad (9)$$

where $p_j$, $m_j$, and $\sigma_j^2$ are volumetric proportion, mean, and variance, respectively. We assume that the means and variances of smaller scale units are such that the assumption (log normality of the global population) in equation (6) is still valid. Equation (9) is an exact equality and requires no further assumptions. The multimodal covariance functions of $\ln K$ and $\ln K_d$ could be found in previous studies as (e.g., Dai et al., 2004; Soltanian et al., 2014; Soltanian et al., in revision):

$$C_Y(\xi) = \sum_{j=1}^{N} p_j^2 \sigma_j^2 e^{-\frac{\xi}{\lambda_j}} + \sum_{i=1}^{N} p_j(1-p_j)\sigma_j^2 e^{-\frac{\xi}{\lambda_\varphi}} + \frac{1}{2}\sum_{i=1}^{N}\sum_{j=1}^{N} p_i p_j (m_i - m_j)^2 e^{-\frac{\xi}{\lambda_I}} \quad (10)$$

where $\lambda_j$ and $\lambda_I$ are the integral scale of the *j*-th RMA unit and the indicator integral scale of the RMAs, respectively; $\lambda_\varphi = \lambda_j \lambda_I /(\lambda_j + \lambda_I)$. Therefore, for multimodal porous media $C_{RR}$ in equation (5) is written as:

$$C_{RR}(\xi) = (\frac{\rho_b}{n} K_d^G)^2 e^{[\sigma_w^2]}(e^{\sum_{j=1}^{N} p_j^2 \sigma_{wj}^2 e^{-\frac{\xi}{\lambda_j}} + \sum_{i=1}^{N} p_j(1-p_j)\sigma_{wj}^2 e^{-\frac{\xi}{\lambda_\varphi}} + \frac{1}{2}\sum_{i=1}^{N}\sum_{j=1}^{N} p_i p_j (m_{wi} - m_{wj})^2 e^{-\frac{\xi}{\lambda_I}}} - 1) \quad (11)$$

where $m_{wj}$, and $\sigma_{wj}^2$ are the mean and variance of $\ln K_d$ of the j-th RMA, respectively.

In order to derive the expression for $C_{v_1 R}$ we use the longitudinal velocity perturbation, $v_1'$, in real space presented by Gelhar and Axness (1983) as:

$$v_1' = \frac{K^G}{n}(Jf - \frac{\partial h}{\partial x_1}) \quad (12)$$

In Fourier space equation (12) could be found using the spectral representation of $\frac{\partial h}{\partial x_1}$ as follows:

$$v_1' = \frac{K^G J}{n}(1 - \frac{k_1^2}{k^2})f \quad (13)$$

where $K^G$ is the geometric mean of the hydraulic conductivity, $f$ is the perturbation of $\ln K$, $h$ is the perturbation of the piezometric head, and $k = (k_1, k_2, k_3)^T$ is a three-dimensional wave-number vector (see also Appendix A in Rajaram, 1997). In order to better explain the derivation of $C_{v_1 R}$ we use equation (12) below.

We intend for our analysis to pertain to media in which the variance of natural-log hydraulic conductivity is less than 1, as is true in the example problem specifically analyzed by Deng et al. (2013). Note that the approximation for $v_1'$ in equation (12) has been shown to work well for the range of variance of hydraulic conductivity <1.0, as considered here (e.g., Bellin et al., 1992; Glimm et al., 1993; Hsu et al., 1996). Highly heterogeneous porous media is outside the scope of this paper.

The product of $R'$ and $v_1'$ is obtained from equation (4) and (12) as:

$$v_1'R' = \frac{\rho_b}{n^2} K^G K_d^G (Jf(e^{w'} - e^{[\frac{\sigma_w^2}{2}]}) - \frac{\partial h}{\partial x_1}(e^{w'} - e^{[\frac{\sigma_w^2}{2}]})) \quad (13)$$

Using the Taylor series expansion for $e^{w'}$ and considering the point that the odd moments of a log normal distribution are zero the term $f(e^{w'} - e^{[\frac{\sigma_w^2}{2}]})$ can be written as:

$$\begin{aligned} f(e^{w'} - e^{[\frac{\sigma_w^2}{2}]}) &= E[f(1 + w' + \frac{w'^2}{2!} + \frac{w'^3}{3!} + ...)] \\ &= E[f + fw' + f\frac{w'^2}{2!} + f\frac{w'^3}{3!} + ..] \\ &= fw'(1 + \frac{1}{3!}\sigma_w^2 + ...) \quad (14) \end{aligned}$$

For the range of variance considered here equation (14) could be well approximated as:

$$fw'(1 + \frac{1}{3!}\sigma_w^2 + ...) = fw'e^{\frac{\sigma_w^2}{3!}} \quad (15)$$

Also, one can find the exact solution of equation (14) using hyperbolic sine function as:

$$fw'(1 + \frac{1}{3!}\sigma_w^2 + ...) = fw'\frac{\sinh(\sigma_w)}{\sigma_w} \quad (16)$$

For the range of variance used in this note the difference between (15) and (16) is negligible. We use equation (16) in our derivation.

Following Rajaram (1997) the cross-spectral density function $S_{v_1 R}(k)$ is obtained from (13) and (15) using the spectral representation of $\dfrac{\partial h}{\partial x_1}$ from Gelhar and Axness (1983) as:

$$S_{v_1 R}(k) = \frac{\rho_b}{n^2} K^G K_d^G J \frac{\sinh(\sigma_w)}{\sigma_w} (1 - \frac{k_1^2}{k^2}) S_{fw}(k) \qquad (17)$$

where $S_{fw}(k)$ is the spectral density of the fluctuations of $f - w$.

Similar to Rajaram (1997) and Deng et al. (2013) $K$ and $K_d$ are assumed to be perfectly correlated as $\ln K_d(x) = a \ln K(x) + b$, where $a$ and $b$ are real constants. Using this model it is easily seen that $S_{fw}(k) = a S_{ff}(k)$. Therefore, the $S_{v_1 R}(k)$ is expressed as:

$$S_{v_1 R}(k) = \frac{\rho_b}{n^2} K^G K_d^G J a \frac{\sinh(\sigma_w)}{\sigma_w} (1 - \frac{k_1^2}{k^2}) S_{ff}(k) \qquad (18)$$

By taking the Fourier transform of equation (18) the cross-covariance function $C_{v_1 R}$ could be found. Of course the resulting cross-covariance function depends on the spectral density function $S_{ff}(k)$. For an isotropic exponential covariance function $C_{ff}(\xi) = \sigma_f^2 e^{-\frac{\xi}{\lambda}}$ the spectral density function $S_{ff}(k)$ is found as:

$$S_{ff}(k) = \frac{\sigma_f^2 \lambda^3}{\pi^2 (1 + \lambda^2 k^2)^2} \qquad (19)$$

For multimodal porous media the corresponding spectral density function for the covariance function in equation (10) is found as (see also Deng et al., 2013):

$$S_{ff}(k) = \sum_{j=1}^{N} p_j^2 \frac{\sigma_{fj}^2 \lambda_j^3}{\pi^2 (1+\lambda_j^2 k^2)^2} + \sum_{i=1}^{N} p_j(1-p_j) \frac{\sigma_{fj}^2 \lambda_\varphi^3}{\pi^2 (1+\lambda_\varphi^2 k^2)^2}$$
$$+ \frac{1}{2}\sum_{i=1}^{N}\sum_{j=1}^{N} p_i p_j (m_{fi} - m_{fj})^2 \frac{\lambda_I^3}{\pi^2 (1+\lambda_I^2 k^2)^2} \qquad (20)$$

where $m_{fj}$, and $\sigma_{fj}^2$ are the mean and variance of $\ln K$ of the $j$-th RMA, respectively. Substituting (20) into (18), using the relationship between the spectrum and the covariance function, and integrating (18) over wave number space, the cross-covariance function $C_{v_1 R}$ for the multimodal isotropic porous medium is found as:

$$C_{v_1 R}(\xi) = \frac{\rho_b}{n^2} K^G K_d^G Ja \frac{\sinh(\sigma_w)}{\sigma_w} \{ \sum_{i=1}^{N} p_i^2 \sigma_{fi}^2 F_1(\frac{\xi}{\lambda_i}) + \sum_{i=1}^{N} p_i p_j \sigma_{fi}^2 F_1(\frac{\xi}{\lambda_\varphi})$$
$$+ \frac{1}{2}\sum_{i=1}^{N}\sum_{j=1}^{N} p_i p_j (m_{fi} - m_{fj})^2 F_1(\frac{\xi}{\lambda_I}) \} \qquad (21)$$

where $F_1(\frac{\xi}{\lambda}) = 4(\frac{\lambda}{\xi})^3 (1-e^{-\frac{\xi}{\lambda}}) - 4(\frac{\lambda}{\xi})^2 e^{-\frac{\xi}{\lambda}} - 2(\frac{\xi}{\lambda})e^{-\frac{\xi}{\lambda}}$. Consequently, the cross-covariance of $v_1$ and $R$ is:

$$\sigma_{v_1 R} = \frac{2}{3}\frac{\rho_b}{n^2} K^G K_d^G Ja \frac{\sinh(\sigma_w)}{\sigma_w} \{ \sum_{i=1}^{N} p_i^2 \sigma_{fi}^2 + \sum_{i=1}^{N} p_i p_j \sigma_{fi}^2 + \frac{1}{2}\sum_{i=1}^{N}\sum_{j=1}^{N} p_i p_j (m_{fi} - m_{fj})^2 \} \qquad (22)$$

The integration method used to attain equations (13), and (14) can be found in Deng et al. (2013). However, for ease of reference we present the integration method in Appendix A.

Note that we used a nonlinear expansion for $\ln K_d$, equation (4), and first-order for $\ln K$, equation (12). For heterogeneity within the range being considered here, Bellin et al. (1993) and Bellin and Rinaldo (1995), have used the same inconsistent expansion in order to analyze the time-dependent dispersion of reactive solutes (see equation (10a) and (17) in Bellin et al., 1993). Importantly, this inconsistent expansion approach and their results were tested against numerical simulations and validated by Bosma et al. (1993). It has been shown that the linear perturbation used in equation (12) results in a good approximation for perturbation in groundwater velocity because the variability in velocity is small compared to the variability in hydraulic conductivity (e.g , Gelhar, 1993; Rubin, 2003). However, this is not the case when approximating the perturbation of the retardation factor. We show in section 3 that the difference between $\sigma_R^2$ resulting from derivation with linear and non-linear perturbations is significant for the range of variance in the sorption distribution coefficient considered by Deng et al. (2013).

## Results and Discussion

We applied the covariance models in form of equations (6), (11), (21), and (22) to the example presented by Deng et al. (2013) (see Table 1). Table 1 presents the parameter values of the three RMAs within a reactive mineral facies (RMF). The global variance of $\ln K_d$ is 0.84 in this example well above the limit of a linear approximation. Using equations (6) and (22) we calculated $\sigma_R^2$ and $\sigma_{v_1 R}$ as 20.25 and 0.228, respectively, whereas Deng et al. (2013), using first-order linear approximation gave $\sigma_R^2$ and $\sigma_{v_1 R}$ as 5.6 and 0.2, respectively. There is a small difference in calculating $\sigma_{v_1 R}$ but in the case of $\sigma_R^2$ the difference is significant. . Therefore, as discussed in the previous section it is important to use a non-linear perturbation for the retardation factor for the range of variance considered here.

In the present note the developed theory is assumed to be valid for the aquifers with small variability ($\sigma_f^2, \sigma_w^2$ <1). For $\sigma_f^2$ and $\sigma_w^2$ larger than unity we suggest that further tests against numerical simulations (e.g., Monte Carlo simulation) could be done to explore the full range of validity with this approach. However, it is outside the scope of this note.

Table 1. Parameters from the example used by Deng et al. (2013).

| RMF | RMA | $L_j$ | $p_j$ | Parameters | $m_j$ | $\sigma_j^2$ | $M_j^G$ | $\lambda_j$ | $\lambda_\varphi$ | $R_j$ |
|---|---|---|---|---|---|---|---|---|---|---|
| Cc-Clay-OM RMF | Cc-QF | 50.0 | 0.6 | ln $K$ | 1.5 | 0.6 | 4.48 | 10 | 6.67 | 2.39 |
|  |  |  |  | ln $K_d$ | -2.2 | 0.22 | 0.11 | 12 | 7.5 |  |
|  | Clay-Fe$_2$O$_3$-QF | 23.5 | 0.15 | ln $K$ | 0.5 | 0.3 | 1.65 | 6 | 4.62 | 4.76 |
|  |  |  |  | ln $K_d$ | -1.2 | 0.12 | 0.3 | 8 | 5.71 |  |
|  | Clay-OM-QF | 26.7 | 0.25 | ln $K$ | 0.05 | 0.15 | 1.05 | 9 | 6.21 | 10.26 |
|  |  |  |  | ln $K_d$ | -0.3 | 0.1 | 0.74 | 7 | 5.19 |  |
| Parameters | ln $K$ |  |  | ln $K_d$ |  |  |  | $R$ |  |  |
| Statistics | $M_Y$ | $\sigma_Y^2$ | $M_Y^G$ | $M_Y$ | $\sigma_Y^2$ | $M_Y^G$ | $M_Y$ | $\sigma_Y^2$ | $M_Y^G$ |  |
| Values | 0.99 | 0.86 | 2.68 | -1.58 | 0.84 | 0.21 | 4.93 | 20.25 | 3.59 |  |
| Parameters | $\bar{v}_1$ | $\sigma_{v_1 R}^2$ |  | $\lambda_I$ | $n$ | $\rho_b$ | $J$ |  |  |  |
| Values | 0.21 | 0.228 |  | 20.0 | 0.2 | 2.5 | 0.01 |  |  |  |

Note: RMA = reactive mineral assemblage within a reactive mineral facies (RMF), Cc = calcite, Fe$_2$O$_3$ = iron oxides, QF = quartz and feldspar; For $j$-th RMA ($j$ = 1, 2, 3), $L_j$ = facies mean length (m), $M_j^G$ = geometric mean, $\lambda_j$ = correlation length (m), $\lambda_\varphi = \lambda_i \lambda_I / (\lambda_i + \lambda_I)$, $R_j$ = retardation factor; $M_Y$ = global mean, $\sigma_Y^2$ = global variance, $M_Y^G$ = global geometric mean, $\bar{v}_1$ = mean flow velocity (m/d), $\sigma_{v_1 R}^2$ = cross-covariance of flow velocity and retardation factor, $\lambda_I$ = indicator correlation length (m), $n$ = porosity, $\rho_b$ = bulk density of the porous media (g/cm$^3$), $J$ = average hydraulic gradient, $K$ = hydraulic conductivity (m/d), $K_d$ = sorption coefficient (cm$^3$/g).

The time-dependent effective $R$ is plotted in Fig. 1 for three cases of correlations between ln $K$ and ln $K_d$: positively correlated ($a$ = 1), uncorrelated ($a$ = 0), and negatively correlated ($a$ = -1). In all three cases, the effective $R$ increases monotonically with time, but effective $R$ starts with different values. The time-dependent effective $R$ in all three cases converges to $\bar{R}$ at the large time limit. As discussed by Rajaram (1997) negative correlation increases the variance which leads to more deviation from the large-time limit (e.g. higher velocity and lower effective $R$). Fig.1 shows that the time-dependent effective $R$ is larger than $\bar{R}$ between times of approximately 800 to 2000 days for the positively correlated case. This behavior was discussed by Rajaram (1997) and attributed to the fact that the positive correlation offsets the influence of spatial variability (see also Garabedian et al., 1988).

There are significant differences from the results presented by Deng et al. (2013). First, the new expressions for $C_{RR}$ and $C_{v_1 R}$ change the starting point and shape of the growth of $\tilde{R}(t)$ (c.f. Deng et al., 2013, Figure 2A). Furthermore, in the results of Deng et al. (2013) the $\tilde{R}(t)$ for the two correlated cases is larger than that for the uncorrelated case, especially at early times.

With the non-linear approximation used here, the $\tilde{R}(t)$ for the non-correlated case properly falls between that for the positive and negatively correlated cases over all time before convergence on the large-time limit.

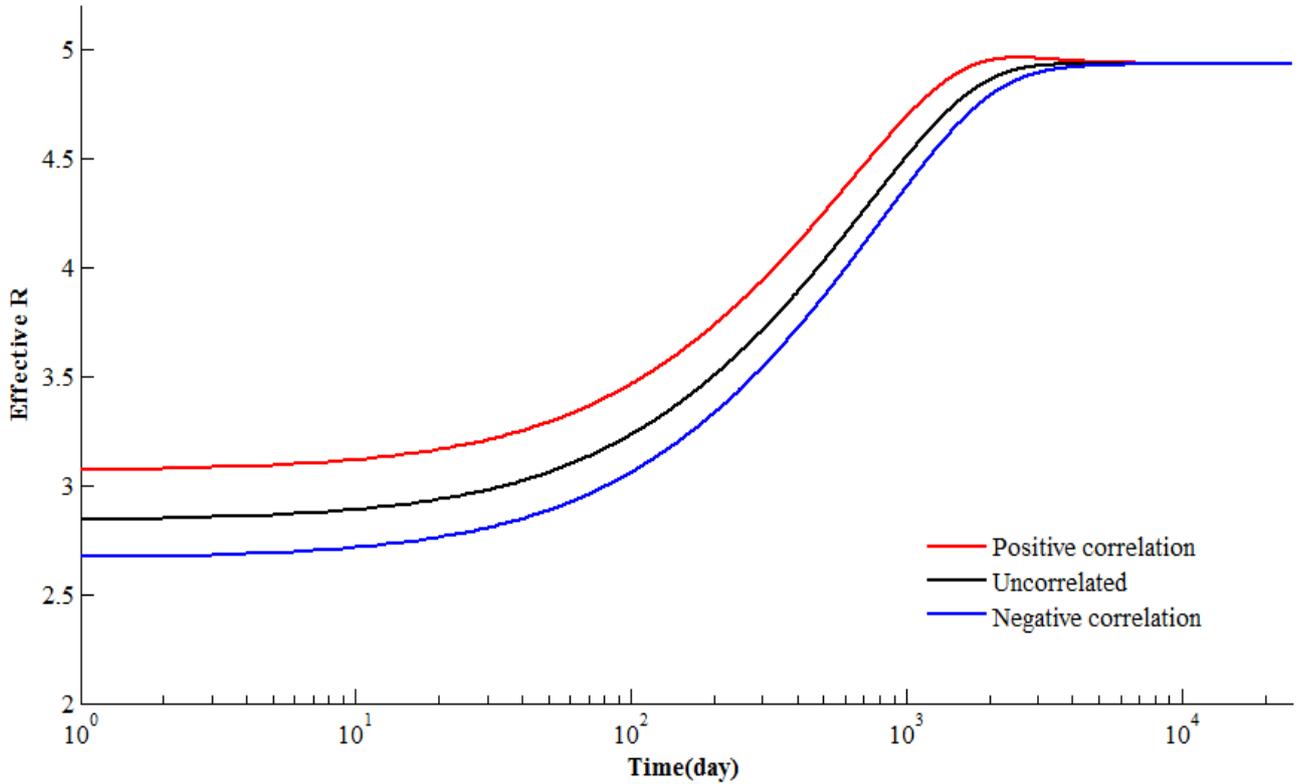

Fig. 1. Time-dependent effective retardation factor calculated using higher-order approximation.

Fig. 2 shows that $\tilde{R}(t)$ changes with the indicator correlation length ($\lambda_I$) when the time is fixed at 1000 $d$. In Fig. 2 the general shape for all three cases is the same. In the case of positive correlation, the $\tilde{R}$ stays constant to a maximum at about 10 m, and then gradually decreases until $\lambda_I$ approximately reaches to 2000 m. This is also true for both the negative correlation and the uncorrelated case. Although $\tilde{R}$ decreases to a minimum for three cases, it reaches to different minimum values. This reveals the influence of the cross-covariance function. Note that in all cases $\tilde{R}$ starts at its value in large time limit as shown in Fig.1 because when $\lambda_I$ is infinitesimal, the full heterogeneity is immediately sampled by the reactive plume.

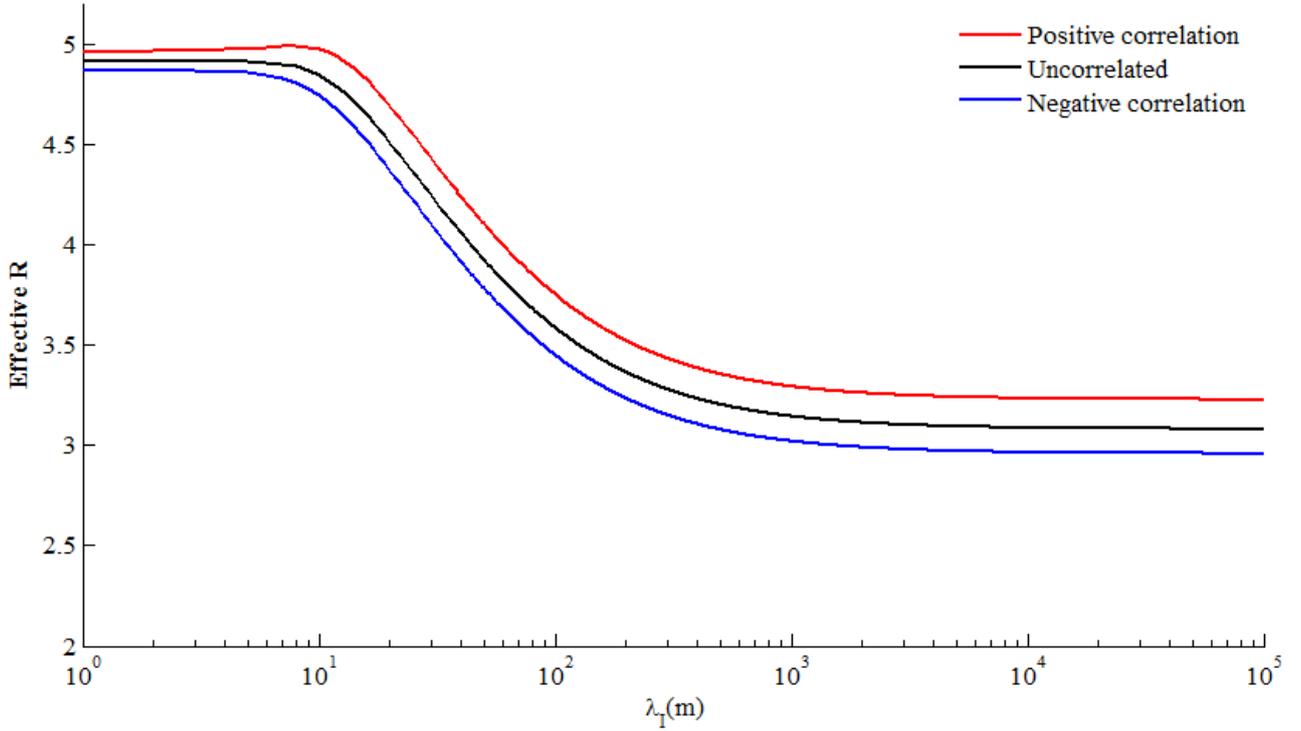

Fig. 2. Time dependent effective retardation factors calculated using higher-order approximation vs. indicator correlation length. Time fixed at 1000 d.

### Acknowledgements

The first author was supported by the National Science Foundation under grant EAR-0810151, and also by a Graduate Fellowship from the College of Science and Mathematics at Wright State University. Any opinions, findings and conclusions or recommendations expressed in this article are those of the authors and do not necessarily reflect those of the National Science Foundation or other supporting institutions. The manuscript was improved based on reviews by Timothy Ginn and six other anonymous reviewers.

### Appendix A. Derivation of cross-covariance of $v_1$ and $R$

The derivation of $C_{v_1 R}(\xi)$ by Deng et al. (2013) is presented here for ease of reference. The $C_{v_1 R}(\xi)$ is found by taking the Fourier transform of the equation (18). Here we consider a unimodal porous media with the spectral density function $S_{ff}(k)$ as in equation (19). Note that the same integration method is used three times for the three exponential terms in equation (20). The $C_{v_1 R}(\xi)$ is found by:

$$C_{v_1 R}(k) = \frac{\rho_b}{n^2} K^G K_d^G Ja \frac{\sinh(\sigma_w)}{\sigma_w} \int\!\!\!\int\!\!\!\int_{-\infty}^{\infty} e^{i\kappa\cdot\xi}(1-\frac{k_1^2}{k^2})\frac{\sigma_f^2 \lambda^3}{\pi^2(1+\lambda^2 k^2)^2} d\kappa \qquad (A1)$$

We set $I$ equal to the integration part in (A1). Thus,

$$I = \int\!\!\!\int\!\!\!\int_{-\infty}^{\infty} e^{ik\cdot\xi}(1-\frac{k_1^2}{k^2})\frac{\sigma_f^2 \lambda^3}{\pi^2(1+\lambda^2 k^2)^2} dk \qquad (A2)$$

One can use a spherical coordinate system and define the following:

$$k_1 = k\cos\beta \qquad (A3)$$
$$k_1/k = \cos\beta\cos\chi + \sin\theta\sin\chi\cos\alpha \qquad (A4)$$
$$dk_1 dk_2 dk_3 = k^2 \sin\theta\, dk\, d\alpha\, d\theta \qquad (A5)$$

where $\chi$ is the angle between the separation vector $\xi$ and the direction of mean flow $k_1$, and $\theta$ is the angle between $\xi$ and $\kappa$. The $\chi$ and $\xi$ are coordinates of the covariance function. The $k$, $\theta$, and $\alpha$ are spherical coordinates in wave number space. Substituting (A3), (A4), and (A5) into (A2) gives:

$$I = \frac{\sigma_f^2}{\lambda\pi^2} \int_{k=0}^{\infty}\int_{\alpha=0}^{2\pi}\int_{\theta=0}^{\pi} \{[1-\cos^2\theta\cos^2\chi - \sin^2\theta\sin^2\chi\cos^2\theta$$
$$-2\cos\theta\cos\chi\sin\theta\sin\chi\cos\alpha]\frac{k^2\lambda^4}{(1+k^2\lambda^2)^2}e^{ik\xi\cos\theta}\sin\theta\, dk\, d\theta\, d\alpha\} \qquad (A6)$$

One can let $\cos\theta = y$ and use the relationship of $e^{ik\xi y} = \cos k\xi y + i\sin k\xi y$ to change the (A6) to the following expression:

$$I = \frac{\sigma_f^2}{\lambda \pi^2} \int_{k=0}^{\infty} \int_{\alpha=0}^{2\pi}$$

$$\times \int_{y=-1}^{1} \{[(1-y^2\cos^2\chi)-(1-y^2)\sin^2\chi\sin^2\alpha]\frac{k^2\lambda^4}{(1+k^2\lambda^2)^2}\cos k\xi y\} \, dk \, dy \, d\alpha \qquad (A7)$$

Deng et al. (2013) integrated (A7) by presenting the following integrals:

$$I_a = \int_0^{\infty} \frac{k^2\lambda^4}{(1+k^2\lambda^2)^2} \cos k\xi y \, dk = \frac{\pi}{4}\lambda(1-\frac{|\xi y|}{\lambda})e^{-\frac{|\xi y|}{\lambda}} \qquad (A8)$$

$$I_b = \int_0^{2\pi} (1-y^2)\sin^2\chi \cos^2\alpha \, d\alpha = \pi(1-y^2)\sin^2\chi \qquad (A9)$$

$$I_c = \int_0^{2\pi} (1-y^2\cos^2\chi) \, d\alpha = 2\pi(1-y^2\cos^2\chi) = \pi(2-2y^2\cos^2\chi) \qquad (A10)$$

Substituting (A8), (A9), and (A10) into (A7) gives:

$$I = \frac{\sigma_f^2}{4} \int_{y=-1}^{1} \{[(2-2y^2\cos^2\chi)-(1-y^2)\sin^2\chi](1-\frac{|\xi y|}{\lambda})e^{-\frac{|\xi y|}{\lambda}}\} dy \qquad (A11)$$

Then, one can expand (A11) as:

$$I = \frac{\sigma_f^2}{2}\{\int_0^1 (1+\cos^2\chi)e^{-\frac{\xi y}{\lambda}} dy + \int_0^1 (1-3\cos^2\chi)y^2 e^{-\frac{\xi y}{\lambda}} dy$$

$$-\int_0^1 \frac{\xi}{\lambda}(1+\cos^2\chi) y e^{-\frac{\xi y}{\lambda}} dy - \int_0^1 \frac{\xi}{\lambda}(1-3\cos^2\chi) y^3 e^{-\frac{\xi y}{\lambda}} dy\}$$

$$= \frac{\sigma_f^2}{2}(I_1 + I_2 - I_3 - I_4) \qquad (A12)$$

Next, one can find $I_1, I_2, I_3$, and $I_4$ as (Deng et al., 2013):

$$I_1 = \int_0^1 (1+\cos^2 \chi) e^{-\frac{\xi y}{\lambda}} dy = -\frac{\lambda}{\xi}(1+\cos^2 \chi)(e^{-\frac{\xi}{\lambda}} - 1) \tag{A13}$$

$$I_2 = \int_0^1 (1-3\cos^2 \chi) y^2 e^{-\frac{\xi y}{\lambda}} dy = (3\cos^2 \chi - 1)[(\frac{\lambda}{\xi})e^{-\frac{\xi}{\lambda}} + 2(\frac{\lambda}{\xi})^2 e^{-\frac{\xi}{\lambda}} + 2(\frac{\lambda}{\xi})^3(e^{-\frac{\xi}{\lambda}} - 1)] \tag{A14}$$

$$I_3 = \int_0^1 \frac{\xi}{\lambda}(1+\cos^2 \chi) y e^{-\frac{\xi y}{\lambda}} dy = -(1+\cos^2 \chi)[e^{-\frac{\xi}{\lambda}} + (\frac{\lambda}{\xi})(e^{-\frac{\xi}{\lambda}} - 1)] \tag{A15}$$

$$I_4 = \int_0^1 \frac{\xi}{\lambda}(1-3\cos^2 \chi) y^3 e^{-\frac{\xi y}{\lambda}} d = (3\cos^2 \chi - 1)[e^{-\frac{\xi}{\lambda}} + 3(\frac{\lambda}{\xi})e^{-\frac{\xi}{\lambda}} + 6(\frac{\lambda}{\xi})^2 e^{-\frac{\xi}{\lambda}} + 6(\frac{\lambda}{\xi})^3(e^{-\frac{\xi}{\lambda}} - 1)] \tag{A16}$$

Substituting (A13), (A14), (A15), and (A16) into (A12), which in turn goes into (A1), finally gives:

$$C_{v_1 R}(\xi) = \frac{\rho_b}{n^2} K^G K_d^G Ja \frac{\sinh(\sigma_w)}{\sigma_w} \frac{\sigma_f^2}{2} \{(1+\cos^2 \chi) e^{-\frac{\xi}{\lambda}} + (1-3\cos^2 \chi)$$
$$\times [e^{-\frac{\xi}{\lambda}} + 2(\frac{\lambda}{\xi})e^{-\frac{\xi}{\lambda}} + 4(\frac{\lambda}{\xi})^2 e^{-\frac{\xi}{\lambda}} + 4(\frac{\lambda}{\xi})^3(e^{-\frac{\xi}{\lambda}} - 1)]\} \tag{A17}$$

Finally, when the separation vector is parallel to the flow direction, $\chi = 0$ and $\cos^2 \chi = 1$, then:

$$C_{v_1 R}(\xi) = \frac{\rho_b}{n^2} K^G K_d^G Ja \frac{\sinh(\sigma_w)}{\sigma_w} \sigma_f^2 [4(\frac{\lambda}{\xi})^3(1 - e^{-\frac{\xi}{\lambda}}) - 4(\frac{\lambda}{\xi})^2 e^{-\frac{\xi}{\lambda}} - 2(\frac{\lambda}{\xi})e^{-\frac{\xi}{\lambda}}] \tag{A18}$$